# Origin of inverse Rashba-Edelstein effect detected at Cu/Bi interface using lateral spin valves

Miren Isasa[1], M. Carmen Martínez-Velarte[2,3], Estitxu Villamor[1], César Magén[2,3,4], Luis Morellón[2,3], José M. De Teresa[2,3,5], M. Ricardo Ibarra[2,3], Giovanni Vignale[6,7], Evgueni V. Chulkov[8,9,10], Eugene E. Krasovskii[8,9,11], Luis E. Hueso[1,11], Fèlix Casanova[1,11,*]

[1]CIC nanoGUNE, 20018 Donostia-San Sebastian, Basque Country, Spain.
[2]Laboratorio de Microscopías Avanzadas (LMA), Instituto de Nanociencia de Aragón (INA), Universidad de Zaragoza, Edificio I+D, 50018 Zaragoza, Spain.
[3]Departamento de Física de la Materia Condensada, Universidad de Zaragoza, 50009 Zaragoza, Spain.
[4]Fundación ARAID, 50004 Zaragoza, Spain.
[5]Instituto de Ciencia de Materiales de Aragón (ICMA), Universidad de Zaragoza-CSIC, Facultad de Ciencias, 50009 Zaragoza, Spain.
[6]Department of Physics and Astronomy, University of Missouri, Columbia, Missouri 65211, USA
[7]Italian Institute of Technology at Sapienza and Dipartimento di Fisica, Università La Sapienza, Piazzale Aldo Moro 5, 00185 Rome, Italy
[8]Departamento de Física de Materiales, Facultad de Ciencias Químicas, Universidad del País Vasco/Euskal Herriko Unibertsitatea, 20080 Donostia-San Sebastian, Basque Country, Spain.
[9]Donostia International Physics Center (DIPC), 20018 Donostia-San Sebastian, Basque Country, Spain.
[10]Centro de Física de Materiales (CFM-MPC) Centro Mixto CSIC-UPV/EHU, 20018 Donostia-San Sebastian, Basque Country, Spain
[11]IKERBASQUE, Basque Foundation for Science, 48013 Bilbao, Basque Country, Spain.

*E-mail: f.casanova@nanogune.eu

The spin transport and spin-to-charge current conversion properties of bismuth are investigated using permalloy/copper/bismuth (Py/Cu/Bi) lateral spin valve structures. The spin current is strongly absorbed at the surface of Bi, leading to ultra-short spin diffusion lengths. A spin-to-charge current conversion is measured, which is attributed to the inverse Rashba-Edelstein effect at the Cu/Bi interface. The spin-current-induced charge current is found to change direction with increasing temperature. A theoretical analysis relates this behavior to the complex spin structure and dispersion of the surface states at the Fermi energy. The understanding of this phenomenon opens novel possibilities to exploit spin-orbit coupling to create, manipulate, and detect spin currents in 2D systems.

Spin-orbit interaction is an essential ingredient in materials and interfaces that has been gaining interest in the last years due to the advantages it offers to exploit the coupling between spin and orbital momentum of electrons in *spintronic* devices [1], leading to the emerging field of *spin-orbitronics* [2]. For instance, magnetization switching of ferromagnetic elements has been recently achieved with torques arising from mechanisms such as spin Hall, Rashba or Dresselhaus effects [3,4]. Of particular interest is the spin Hall effect (SHE), which can be used to create and detect a pure spin current without the use of ferromagnets or magnetic fields. This is a phenomenon appearing in materials with strong spin-orbit coupling (SOC), in which a charge current flowing through a non-magnetic material creates a spin current in the transverse direction to the charge current [5,6]. Reciprocally, a spin current through a non-magnetic material creates a transverse charge current, *i.e.*, the inverse SHE (ISHE) [7-9]. Very recently, a new way of converting spin current into charge current has been experimentally reported: the inverse Rashba-Edelstein effect (IREE) [10,11]. This phenomenon arises from the spin-orbit splitting in a two-dimensional electron gas (2DEG) known as the Rashba effect (Figure 1(a)), leading to the conversion of a 3D spin current into a 2D charge current [12]. There are many systems where the surface state is strongly spin-orbit split, including metals (a typical example is Au(111), Ref. 13) and semiconductors with giant SOC, BiTeI and BiTeCl [14,15], although in these cases the bulk states usually dominate the conduction. An optimal choice seems to be a semimetallic system such as bismuth.

Bismuth in particular is a group V semimetal with an anisotropic Fermi surface, where small electron and hole pockets give rise to a low carrier density, $n \sim p \sim 3 \cdot 10^{17}$ cm$^{-3}$, high resistivity (~100 μΩ·cm) and relatively large Fermi wavelength (~30 nm) [16]. For thin films, the energy band structure changes. When film dimensions are comparable to the Fermi wavelength, a semimetal-to-semiconductor transition is predicted [17]. At the same time, metallic surface states are found to gain relevance in transport, leading to a 2D confinement of the carriers as recently observed experimentally [16]. The strong SOC in Bi and the loss of inversion symmetry at the surface produces Rashba splitting of the surface states [18]. For this reason, not only the SOC on the Bi surface has attracted a great deal of attention [19], but also surface alloying of Bi with other materials has been studied. The largest spin-splitting has been found for a silver (Ag)/Bi interface [20,21], however other systems such as copper (Cu)/Bi are also expected to manifest a sizeable effect [22].

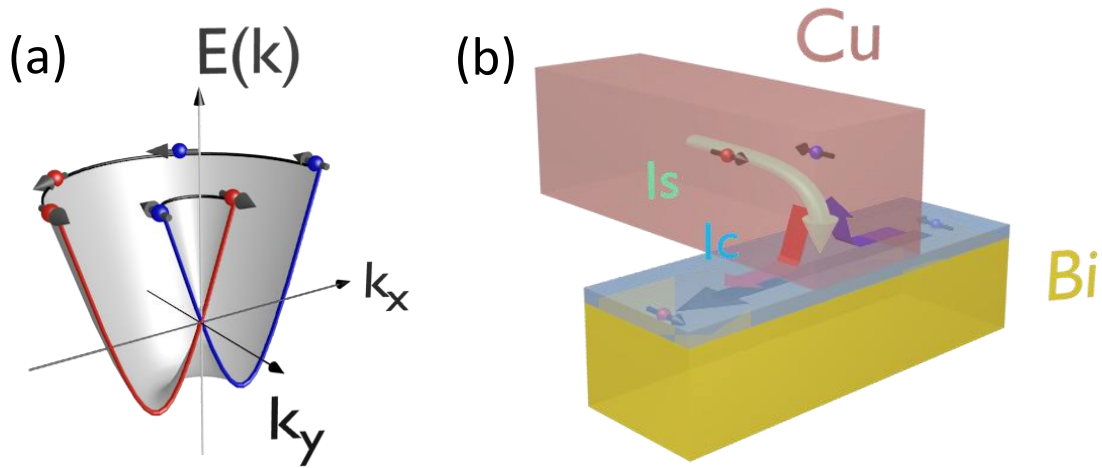

FIG 1. (a) Rashba energy dispersion for a 2D electron gas (2DEG). (b) Schematic representation of the detection of the inverse Rashba-Edelstein effect using the spin absorption method, where Cu acts as a pure spin current channel. The red (blue) arrows represent the current for spin-up (spin-down) electrons, the green arrow represents the 3D spin current (Is) through the Cu and the black arrow represent the 2D charge current (Ic) through the Bi metallic surface.

In this work, we study the spin transport properties and spin-to-charge conversion in Bi using a lateral spin valve (LSV) structure (Figure 1(b)). By applying the spin absorption method [8,23-27], we observe that Bi strongly absorbs the spin current and demonstrate a spin-to-charge current conversion in the LSV. The analysis of the obtained results leads us to argue that the spin absorption and subsequent spin-to-charge conversion do not occur at the bulk of Bi but at the Cu/Bi interface, therefore detecting IREE. Moreover, we evaluate the IREE length, which characterizes the spin-to-charge conversion ratio, as a function of temperature. This ratio exhibits a sign change at a certain temperature threshold (~125 K). In order to understand this puzzling behavior, we perform a theoretical analysis based on the first-principles band structure, which reveals that the strong spin-splitting of the surface states in Bi (111) is responsible of the IREE and that the non-monotonic dispersion of such states can account for the sign change.

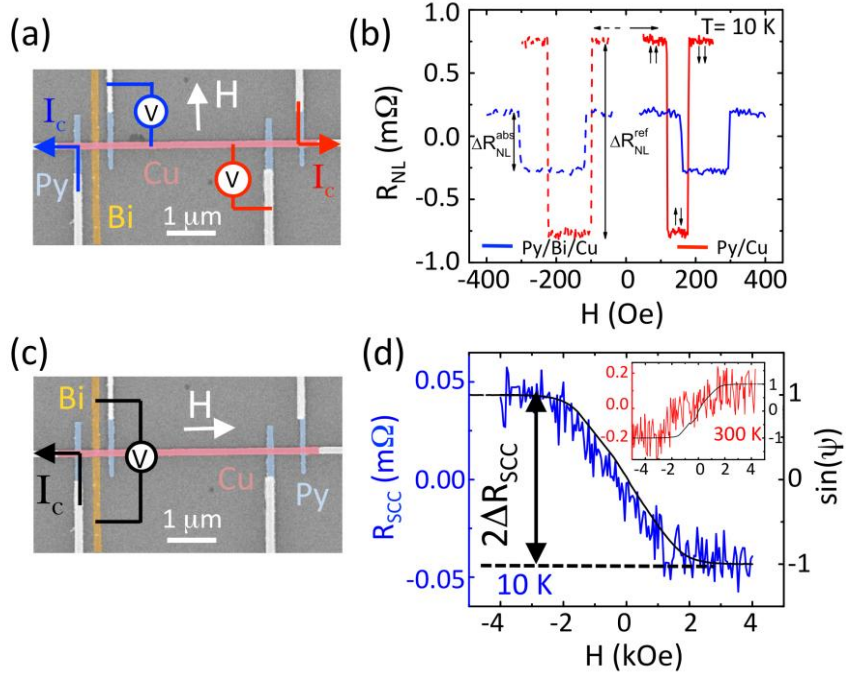

FIG. 2. (a) Colored SEM image of two Py/Cu LSVs, the left one with a Bi wire in between the Py electrodes and the right one without. The measurement configuration, the direction of the applied magnetic field ($H$) and the materials are shown. (b) Red (blue) curve represents the non-local resistance $R_{NL}$ as a function of $H$ at 10 K in a Py/Cu LSV without (with) a Bi wire in between the electrodes. A current of $I_c$=0.1 mA is injected. The corresponding spin signals are tagged as $\Delta R_{NL}^{ref}$ and $\Delta R_{NL}^{abs}$, respectively. The solid (dashed) line represents the increasing (decreasing) sweep of $H$. (c) Colored SEM image of a typical device to measure the spin-to-charge conversion. The materials (Py, Cu and Bi), the direction of the magnetic field ($H$) and the measurement configuration are shown. (d) Non-local resistance $R_{SCC}$ for Bi detected when measuring in the spin-to-charge conversion configuration as a function of $H$ at 10 K (blue curve). The spin-to-charge conversion signal is tagged as $2\Delta R_{SCC}$. A current of $I_c$=1 mA is injected. Inset: $R_{SCC}$ as a function of $H$ at 300 K (red curve). Black solid lines are the sine of the magnetization rotation angle ($\psi$), which serves as a guide for the expected shape of the spin-to-charge conversion curve [32].

We fabricated four samples by multiple-step electron-beam lithography on top of a SiO$_2$ (150 nm)/Si substrate, followed by metal deposition and lift-off. These samples consist of two Cu/permalloy (Py) LSVs, each one with the same separation ($L$~630 nm) in between Py electrodes. The only difference between both LSVs is that one of them has an additional Bi wire in between the electrodes (see Fig. 2(a)). The two pairs of Py electrodes were patterned in the first lithography step and 35 nm of Py were e-beam evaporated. Different widths of Py electrodes were chosen, ~95 nm and ~130 nm, in order to obtain different switching magnetic fields. In the second lithography step, the middle wire in between one of the two pairs of electrodes was patterned. Afterwards, ~150-nm-wide and 20-nm-thick Bi was e-beam evaporated at a pressure of ~1×10$^{-7}$ mbar. Since our Bi films grow on top of SiO$_2$, they are predominantly textured along the (111) direction [28]. In the third lithography step, the ~150-nm-wide channel was patterned and 100-nm-thick Cu was thermally evaporated at a pressure of ≤ 1×10$^{-8}$ mbar. Before the Cu deposition, the Py and Bi wire surfaces were cleaned by Ar-ion milling to remove the possible resist leftovers and oxide formation. Figure 2(a) is a scanning electron microscope (SEM) image of a sample showing the two pairs of LSVs, with (left LSV) and without (right LSV) Bi wire. Although the measurements for all four samples yield similar results, for the sake of simplicity we will mostly show the results obtained for one of them (sample 1).

Non-local transport measurements have been carried out in a liquid-He cryostat (applying an external magnetic field $H$ and varying the temperature) using a "DC reversal" technique [29]. When a charge current $I_c$ is injected from the Py electrode, a spin accumulation is built at the Py/Cu interface that diffuses away along the Cu wire creating a spin current. When it reaches the second Py electrode, a spin accumulation is built at the Cu/Py interface, which will result in a measurable voltage, $V$. This $V$ normalized to the injected current, $I_c$, is defined as the non-local resistance $R_{NL}=V/I_c$ (see Fig. 2(a) for a measurement scheme). $R_{NL}$ changes from positive to negative when the relative magnetization of the Py electrodes changes from a parallel to an antiparallel state by sweeping $H$. The change in $R_{NL}$ is defined as the spin signal, $\Delta R_{NL}^{ref}$, which is proportional to the spin accumulation at the detector (red curve in Fig. 2(b)). If a middle wire (Bi in this case) is inserted in between the Py electrodes, spin absorption into the Bi occurs, and thus the detected spin signal, $\Delta R_{NL}^{abs}$, decreases (blue curve in Fig. 2(b)). By normalizing the two different spin signals ($\Delta R_{NL}^{ref}$ and $\Delta R_{NL}^{abs}$) we can define the parameter $\eta$, which is related to the efficiency of the Bi wire to absorb the spin current diffusing along the Cu channel. The one-dimensional spin diffusion model gives us a relation between $\eta$ and the spin diffusion length of the middle wire through the following equation [8,23]:

$$\eta = \frac{\Delta R_{NL}^{abs}}{\Delta R_{NL}^{ref}} = \frac{R_{Bi}\sinh(L/\lambda_{Cu})+R_{Bi}\frac{R_{Py}}{R_{Cu}}e^{(L/\lambda_{Cu})}+\frac{R_{TM}}{2}\left(\frac{R_{Py}}{R_{Cu}}\right)^2 e^{(L/\lambda_{Cu})}}{R_{Cu}(\cosh(L/\lambda_{Cu})-1)+R_{Bi}\sinh(L/\lambda_{Cu})+R_{Py}\left[e^{(L/\lambda_{Cu})}\left(1+\frac{R_{Py}}{2R_{Cu}}\right)\left(1+\frac{R_{Bi}}{R_{Cu}}\right)-1\right]}, \quad (1)$$

where $R_{Bi} = \frac{\rho_{Bi}\lambda_{Bi}}{w_{Bi}w_{Cu}\tanh(t_{Bi}/\lambda_{Bi})}$, $R_{Cu} = \frac{\rho_{Cu}\lambda_{Cu}}{2w_{Cu}t_{Cu}}$ and $R_{Py} = \frac{\rho_{Py}\lambda_{Py}}{(1-\alpha_{Py}^2)w_{Cu}w_{Py}}$ are the spin resistances, $\lambda_{Bi,Cu,Py}$ the spin diffusion lengths, $\rho_{Bi,Cu,Py}$ resistivities, $w_{Bi,Cu,Py}$ widths and $t_{Bi,Cu}$ thicknesses of the Bi, Cu and Py, respectively. $\alpha_{Py}$ is the current spin polarization of Py. Since $\lambda_{Cu}$, $\alpha_{Py}$ and $R_{Cu,Py}$ values are well known from our previous work [30,31], all the geometrical parameters are measured by SEM and $\rho_{Bi}$ is measured in the same device in which the spin signals are measured [32], $\lambda_{Bi}$ can be directly obtained from Eq. (1). This spin absorption (SA) technique has been successfully used to measure short spin diffusion lengths in metals before [8,23-27].

From our experiments at low temperature, we obtain a spin absorption ratio of $\eta$= 0.140 ± 0.008, which together with the measured $\rho_{Bi}$=988 μΩ cm at 10 K, yields

$\lambda_{Bi}$=0.050 ± 0.005 nm. However, this value is far from $\lambda_{Bi}$=20 nm obtained by weak antilocalization (WAL) measurements in Bi evaporated under the same conditions [32,33,34,35]. The same occurs at room temperature, where from the measured values of $\eta$=0.11± 0.04 and $\rho_{Bi}$=830 μΩ cm we extract a spin diffusion length of $\lambda_{Bi}$=0.11± 0.05 nm. This value is again far from room temperature $\lambda_{Bi}$ values reported in literature using spin-pumping technique, which range from 8 to 50 nm [11,36,37]. We must stress here that WAL and spin-pumping experiments probe the bulk $\lambda_{Bi}$ value. However, both the room- and low-temperature $\lambda_{Bi}$ values that we extract from SA measurements [32] are anomalously small, as they are shorter than the interatomic distance of Bi [38], evidencing that the spin current is strongly absorbed at the metallic surface rather than in the bulk, in good agreement with the unique surface properties of Bi [16].

Once this spin current absorption is confirmed, we can now study the spin-to-charge current conversion (SCC) in the Cu/Bi interface. For this experiment we use the same device in which SA is measured with the configuration shown in Fig. 2(c). Using the Py electrode as a spin current injector, a three-dimensional (3D) spin current is created along the Cu channel, which will be partially absorbed into the Bi wire. The ratio between the injected charge current, $I_c$, and the spin current reaching the Bi wire, $I_s$, is defined as [8,23]:

$$\frac{I_s}{I_c} = \frac{\alpha_{Py} R_{Py}\left[\sinh(L/2\lambda_{Cu}) + \frac{R_{Py}}{2R_{Cu}} e^{L/2\lambda_{Cu}}\right]}{R_{Cu}[\cosh(L/\lambda_{Cu})-1] + R_{Py}\left[e^{L/\lambda_{Cu}}\left(1+\frac{R_{Py}}{2R_{Cu}}\right)\left(1+\frac{R_{Bi}}{R_{Cu}}\right)-1\right] + R_{Bi}\sinh(L/\lambda_{Cu})} \quad . \quad (2)$$

This spin current, $I_s$, that is absorbed into the metallic-Bi surface will be converted into a two-dimensional (2D) charge current at the Cu/Bi interface via the inverse Rashba-Edelstein effect (Figure 1(b)), as recently reported for a similar (Ag/Bi) interface [10,11]. The parameter that relates the 3D spin current to the 2D charge current, and therefore quantifies the IREE, is the IREE length, $\lambda_{IREE}$. Although it has length units, $\lambda_{IREE}$ is actually not a physical length. It can be calculated as:

$$\lambda_{IREE} = \frac{t_{Bi}}{\rho_{Bi}} \frac{w_{Bi}}{x} \left(\frac{I_c}{I_s}\right) \Delta R_{SCC,} \quad (3)$$

where $x$ is a correction factor that takes into account the current that is shunted through the Cu, due to its lower resistivity compared to Bi. $x$ is obtained from numerical calculations using a finite elements method (SpinFlow3D software) [25,32]. $\Delta R_{SCC}$ is the change in non-local resistance ($R_{SCC}$) that we measure when a magnetic field is applied in the configuration shown in Fig. 2(c). As can be seen in Fig. 2(d), by increasing the magnetic field, $R_{SCC}$ changes continuously following the magnetization of the Py electrode until it gets saturated above the saturation field [8,23-27]. 2$\Delta R_{SCC}$ is the change in $R_{SCC}$ in between the two saturated regions.

The $\lambda_{IREE}$ value that we extract from our measurements (Fig. 2(d)) is $\lambda_{IREE}$=0.009 ± 0.002 nm (-0.0010 ± 0.0003 nm) at 300 K (10 K), which is smaller than $\lambda_{IREE}$=0.3 nm reported for Ag/Bi at 300 K [10]. It is worth noting that the other measured samples (samples 2, 3 and 4) give similar values, showing the reproducibility of the effect (Fig. 3(a)). Since the injection process might be substantially less efficient for electrical spin injection than for spin pumping experiments [39], our effective $\lambda_{IREE}$ value is a lower limit. A theoretical estimation of $\lambda_{IREE}$ from the expression $\lambda_{IREE}=\alpha_R \tau/\hbar$ [10,12], where $\tau$ is the momentum relaxation time as discussed by Shen *et al.* [12] and $\alpha_R$ is the Rashba coefficient, is not trivial. $\alpha_R$ and $\tau$ can certainly change a lot from a Ag/Bi system to a Cu/Bi system. On the one hand, $\tau$ should be the momentum relaxation time of the metallic surface of Bi, which is not straightforward to determine from experiments, as usually bulk $\tau$ is measured. On the other hand, for the complex non-monotonic dispersion of the Bi(111) surface states (Fig. 4(a)) the parameter $\alpha_R$ does not have an obvious physical meaning, and it is not clear which value should be ascribed to it in the present experiment. Taking anyway $\alpha_R$ =0.56 eV/Å as in Ref. [10], the momentum relaxation time in our Cu/Bi system is calculated to be $\tau = 2\times 10^{-16}$ s, which is about an order of magnitude smaller than the momentum relaxation time estimated in Ref. [10] ($\tau = 5\times 10^{-15}$ s). Although it differs by an order of magnitude, it is consistent with the electrons being in the Dyakonov-Perel diffusive regime, which underlies the calculation of Ref. [12].

One could argue that the ISHE, and not the IREE, is the responsible mechanism to convert spin current into charge current. This would be the case if the spin current diffusing along the Cu channel were absorbed by the bulk Bi, instead of the interface. In such scenario, however, the spin diffusion length obtained from the SA experiment should be much longer, similar to the lengths obtained from WAL [32, 34, 35] or spin pumping [11,36,37] measurements. Since this is not the case, the observed discrepancy could only be compatible with spin absorption in the bulk Bi by assuming a large spin-flip scattering at the interface, leading to spin memory loss (SML) [40]. The ratio between the spin current absorbed into the Bi wire and the total spin current coming from the Cu channel ($r_{SML}$) can be calculated from the SA-measured and the bulk $\lambda_{Bi}$ values [32]. $r_{SML}$ must be taken into account when evaluating the ISHE. The spin Hall angle, $\theta_{SH}$, which quantifies the spin-to-charge current conversion due to ISHE in the bulk, is then calculated to be $|\theta_{SH}|>100\%$ both at 10 and 300 K [32]. This unphysical value rules out the possibility of ISHE as the spin-to-charge current conversion mechanism in our system.

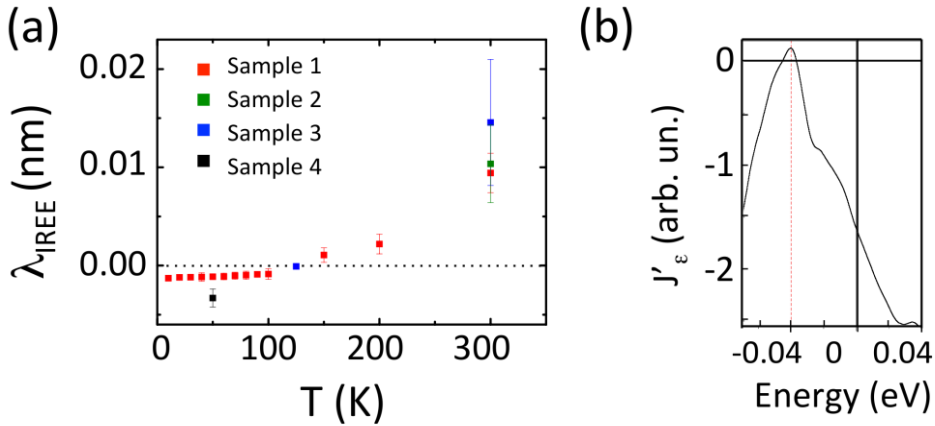

FIG. 3. (a) Temperature evolution of the IREE length of Bi as obtained from four different samples. (b) Energy dependence of the spectral current density $j'_E(E)$ calculated by using $s_\lambda(\mathbf{k}_\parallel)$ for $n_\lambda(\mathbf{k}_\parallel)$.

Once we have determined the mechanism that converts spin current into charge current, we investigate the temperature dependence of the IREE. As can be seen in Fig.

3(a), there is a change in the sign of $\lambda_{IREE}$ between 100 and 150 K. This implies that opposite charge currents are created with the same spin current polarization at low and high temperatures. The $\lambda_{IREE}$ values obtained from samples 2, 3 and 4 confirm that the sign change is very robust.

In order to understand this behavior, a careful microscopic analysis of the spin-resolved surface electronic structure is needed. Let us consider the non-equilibrium distribution of carriers in Bi produced by the injection of a pure spin current. The non-equilibrium carriers are restricted to a close vicinity of the Fermi energy, and the probability of an electron state to host the injected electron depends on its probability to have the respective spin, ↑ or ↓, in the vicinity of the surface (by controlling the overlap between the wave function of the injected electron and the current carrying state).

Following the experimental configuration (Fig. 1(b)), let the in-plane spin quantization axis be perpendicular to the induced current direction and consider the difference between the current due to spin-↑ and spin-↓ electrons. In a semi-infinite crystal, the eigenstates are labeled by the Bloch vector parallel to the surface $\mathbf{k}_\parallel$, the energy $E$, and the band number $\lambda$. In a slab calculation, the energy continuum at each $\mathbf{k}_\parallel$ is approximated by a discrete set of levels. Each eigenstate is characterized by a spin value $s_\lambda(\mathbf{k}_\parallel)$, which is defined as an integral over a surface region from depth $z_0$ to vacuum $z_V$:

$$s_\lambda(k_\parallel) = \int_{z_0}^{z_v} \rho_\lambda^\uparrow(k_\parallel, z) - \rho_\lambda^\downarrow(k_\parallel, z) dz. \tag{4}$$

The spin spectral density for $\mathbf{k}_\parallel$ along $\overline{\Gamma M}$ and the spin quantization axis perpendicular to $\mathbf{k}_\parallel$ are shown in Fig. 4(a) (the integration in Eq. (4) is over the outermost bilayer). The electric current density $\mathbf{j}$ is then a sum of the partial currents over all states outside the equilibrium distribution. The contribution of a narrow energy interval $\delta E$ around energy $E$ to the non-equilibrium current is $\delta j = j'_E(E) \, \delta E$, with the current spectral density given by the integral over a constant energy contour:

$$j'_E(E) = \sum_\lambda \int_{FS} d\tau \frac{n_\lambda(k_\parallel^\lambda) v_\lambda(k_\parallel^\lambda)}{\left|\nabla_{k_\parallel} \epsilon_\lambda(k_\parallel^\lambda)\right|}, \tag{5}$$

where $\mathbf{v}_\lambda(\mathbf{k}_\parallel)$ is the group velocity, and $n_\lambda(\mathbf{k}_\parallel)$ is the deviation of the occupation number from its equilibrium value. At elevated temperatures the Fermi distribution smears out, and the states below $E_F$ become available to the injected electrons, which changes the balance of different contributions to the integral and, thus, may change the sign of the effect (see Fig. 3(b)).

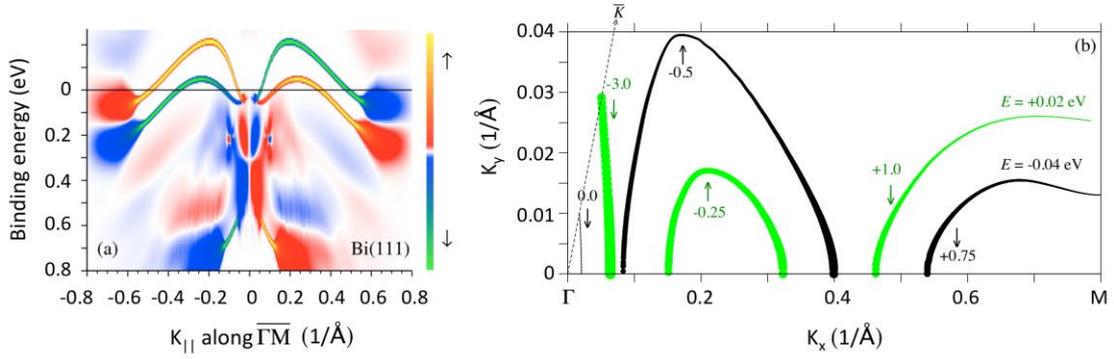

FIG. 4. (a) $\mathbf{k}_\parallel$-projected spin spectral density for $\mathbf{k}_\parallel$ along $\overline{\Gamma M}$. The calculation is performed for a slab of 16 Bi bilayers with the full-potential linear augmented plane wave method [41]. Surface states are shown by distinct thick lines, and the bulk states are presented by smearing the slab levels with a Gaussian of 0.15 eV FWHM. (b) Constant energy contours in the 30° sector of the 2D Brillouin zone for $E = 0.02$ eV (green lines) and $E = -0.04$ eV (black lines) relative to the Fermi energy. The sign of the spin projection $s_\lambda(\mathbf{k}_\parallel)$ of the contour is indicated by ↑ or ↓, and the line thickness is proportional to the absolute value of the spin projection. The value at the ↑ or ↓ symbol (in arbitrary units) indicates the contribution of that branch of the contour to $j'_E$.

Let us consider current along the Bi(111) surface in the $\overline{\Gamma M}$ direction. Because the coefficients $n_\lambda(\mathbf{k}_\parallel)$ are not known (they depend on specific features of the injection process), for a qualitative discussion let us assume $n_\lambda(\mathbf{k}_\parallel)$ to be proportional to the spin at the surface $s_\lambda(\mathbf{k}_\parallel)$, see Eq. (4). Two constant energy contours for two energies close to $E_F$ are shown in Fig. 4(b). Although the bulk states at the Fermi level are spin polarized at the surface, see Fig. 4(a), the main contribution to the inverse Rashba-Edelstein effect turns out to come from the surface states. Within the same surface state band the net spin projection does not change sign, but the direction of the group velocity changes. As a result, the contributions from different $\mathbf{k}_\parallel$ regions have different sign, and their relative weights vary with energy. The function $j'_E(E)$ calculated for a 16-bilayer Bi(111) slab is shown in Fig. 3(b). The $j'_E(E)$ curve turns out to be non-monotonic, and it changes sign at 0.04 eV below the Fermi energy.

This offers the following hypothetical scenario of the sign change in the inverse Rashba-Edelstein effect with increasing the temperature: suppose that in the actual case the current spectral density changes sign just below the Fermi level. As the equilibrium occupation of the states below $E_F$ decreases, they become selectively (depending on the spin) occupied by the injected electrons and may produce a current in the opposite direction. This may not happen for surface states of the Rashba model because of their monotonic dispersion (unless $n_\lambda(\mathbf{k}_\parallel)$ show sharp variations), but this may happen for the more complicated surface states of Bi(111). The present calculation suggests a minor role of the bulk states in IREE, which stems from their low density at $E_F$ (semimetallic character of Bi). Moreover, both the polarization and the group velocity have the same sign for the bulk hole pocket at $\overline{\Gamma}$ and electron pocket at $\overline{M}$, so a change of their occupation numbers does not explain the inversion of the induced current. In spite of the limitations of the present analysis (that arise from our lack of knowledge of the actual structure of the Cu/Bi interface and its $\mathbf{k}_\parallel$- and spin-resolved transport properties), it suggests a microscopic mechanism of converting spin current into charge current via surface states, which possesses the property of changing the sign depending on occupation numbers.

In summary, we demonstrate that the Bi metallic surface acts as a strong spin absorber. We show that a conversion of 3D spin currents to 2D charge currents occurs at such metallic surface by means of the inverse Rashba-Edelstein effect. Moreover, the temperature dependence of the IREE features a sign crossover at ~125 K, which according to our theoretical analysis, arises from a spin structure with non-monotonic dispersion of the surface states at the Fermi level. This rich phenomenology of the complex electronic behavior of Bi could be further exploited to unveil yet unpredicted spin-dependent effects.

**Acknowledgments**

We thank Prof. Albert Fert for fruitful discussions on the experimental results, Omori Yasutomo and Prof. YoshiChika Otani for the calculation of the shunting factor using SpinFlow3D and Dr. Soraya Sangiao for the discussions on the properties of bismuth. This work is supported by the European Commission under the Marie Curie Actions (256470-ITAMOSCINOM), the SUDOE (TRAIN2-SOE2/P1/E-280) and the European Research Council (257654-SPINTROS), by the Spanish MINECO under Projects No. MAT2012-37638, MAT2011-13099-E, MAT2011- 27553-C02, and FIS2013-48286-C2-1-P, including FEDER funds, by the Basque Government under Project No. PI2011-1 and by the Aragon Regional Government under Project No. E26. M.I. and E.V. thank the Basque Government for a PhD fellowship (Grants No. BFI-2011-106 and No. BFI-2010-163).

**REFERENCES**

[1] *Spin Current*, edited by S. Maekawa, S. O. Valenzuela, E. Saitoh, and T. Kimura (Oxford University Press, 2012).
[2] T. Kuschel and G. Reiss, Nature Nanotech.**10**, 22 (2015).
[3] I. M. Miron, K. Garello, G. Gaudin, P.-J. Zermatten, M. V. Costache, S. Auffret, S. Bandiera, B. Rodmacq, A. Schuhl, and P. Gambardella, Nature **476**, 189 (2011); K. Garello, I. M. Miron, C. O. Avci, F. Freimuth, Y. Mokrousov, S. Blügel, S. Auffret, O. Boulle, G. Gaudin, and P. Gambardella, Nature Nanotech. **8**, 587 (2013).
[4] L. Liu, C.-F. Pai, Y. Li, H. W. Tseng, D. C. Ralph, and R. A. Buhrman, Science **336**, 555 (2012).
[5] J. E. Hirsch, Phys. Rev. Lett. **83**, 1834 (1999).
[6] A. Hoffmann, IEEE Trans. Magn. **49**, 5172 (2013).
[7] S. O. Valenzuela and M. Tinkham, Nature **442**, 176 (2006).
[8] M. Morota, Y. Niimi, K. Ohnishi, D. H. Wei, T. Tanaka, H. Kontani, T. Kimura and Y. Otani, Phys. Rev. B **83**, 174405 (2011).
[9] T. Seki, Y. Hasegawa, S. Mitani, S. Takahashi, H. Imamura, S. Maekawa, J. Nitta and K. Takanashi, Nature Mat. **7**, 125 (2008); B. Gu, I. Sugai, T. Ziman, G. Y. Guo, N. Nagaosa, T. Seki, K. Takanashi, and S. Maekawa, Phys. Rev. Lett. **105**, 216401 (2010).
[10] J. C. Rojas-Sanchez, L. Vila, G. Desfonds, S. Gambarelli, J. P. Attané, J.M. De Teresa, C. Magén and A. Fert, Nature Comms. **4**, 2944 (2013).
[11] S. Sangiao, J. M. De Teresa, L. Morellon, I. Lucas, M. C. Martinez-Velarte, and M. Viret, Appl. Phys. Lett. **106**, 172403 (2015).
[12] K. Shen, G. Vignale and R. Raimondi, Phys. Rev. Lett. **112**, 096601 (2014).
[13] S. LaShell, B. A. McDougall, and E. Jensen, Phys. Rev. Lett. **77**, 3419 (1996).

[14] G. Landolt, S. V. Eremeev, Y. M. Koroteev, B. Slomski, S. Muff, T. Neupert, M. Kobayashi, V. N. Strocov, T. Schmitt, Z. S. Aliev, M. B. Babanly, I. R. Amiraslanov, E. V. Chulkov, J. Osterwalder, and J. H. Dil, Phys. Rev. Lett. **109**, 116403 (2012).
[15] S. V. Eremeev, I. A. Nechaev, Yu. M. Koroteev, P. M. Echenique, and E. V. Chulkov, Phys. Rev. Lett. **108**, 246802 (2012).
[16] S. Xiao, D. Wei and X. Jin, Phys. Rev. Lett. **109**, 166805 (2012).
[17] V. N. Lutskii, JETP Lett. **2**, 245 (1965).
[18] J. I. Pascual, G. Bihlmayer, Y. M. Koroteev, H. P. Rust, G. Ceballos, M. Hansmann, K. Horn, E. V. Chulkov, S. Blügel, P. M. Echenique and Ph. Hofmann, Phys. Rev. Lett. **93**, 196802 (2004).
[19] Y. M. Koroteev, G. Bihlmayer, J. E. Gayone, E. V. Chulkov, S. Blügel, P. M. Echenique and Ph. Hofmann, Phys. Rev. Lett. **93**, 046403 (2004).
[20] C. R. Ast, J. Henk, A. Ernst, L. Moreschini, M. C. Falub, D. Pacilé, P. Bruno, K. Kern and M. Grioni, Phys. Rev. Lett. **98**, 186807 (2007).
[21] G. Bihlmayer, S. Blügel, and E. V. Chulkov, Phys. Rev. B **75**, 195414 (2007).
[22] H. Bentmann, F. Forster, G. Bihlmayer, E. V. Chulkov, L. Moreschini, M. Grioni and F. Reinert, EPL **87**, 37003 (2009).
[23] M. Isasa, E. Villamor, L. E. Hueso, M. Gradhand and F. Casanova, Phys. Rev. B **91**, 024402 (2015).
[24] Y. Niimi, M. Morota, D. H. Wei, C. Deranlot, M. Basletic, A. Hamzic, A. Fert and Y. Otani, Phys. Rev. Lett. **106**, 126601 (2011).
[25] Y. Niimi, Y. Kawanishi, D. H. Wei, C. Deranlot, H. X. Yang, M. Chshiev, T. Valet, A. Fert and Y. Otani, Phys. Rev. Lett. **109**, 156602 (2012).
[26] K. Fujiwara, Y. Fukuma, J. Matsuno, H. Idzuchi, Y. Niimi, Y. Otani and H. Takagi, Nature Comms. **4**, 2893 (2013).
[27] Y. Niimi, D. Wei, H. Idzuchi, T. Wakamura, T. Kato, and Y. C. Otani, Phys. Rev. Lett. **110**, 016805 (2013).
[28] N. Marcano, S. Sangiao, C. Magén, L. Morellón, M. R. Ibarra, M. Plaza, L. Pérez, and J. M. De Teresa, Phys. Rev. B **82**, 125326 (2010).
[29] F. Casanova, A. Sharoni, M. Erekhinsky and I. K. Schuller, Phys. Rev. B **79**, 184415 (2009).
[30] E. Villamor, M. Isasa, L. E. Hueso and F. Casanova, Phys. Rev. B **87**, 094417 (2013).
[31] E. Villamor, M. Isasa, L. E. Hueso and F. Casanova, Phys. Rev. B **88**, 184411 (2013).
[32] See Supplemental Material at [*URL will be inserted by publisher*] for further experiments to characterize the Py/Cu/Bi lateral spin valves and for the calculation of the ISHE in Bi by considering spin memory loss at the Cu/Bi interface.
[33] S. Sangiao, N. Marcano, J. Fan, L. Morellón, M. R. Ibarra, J. M. De Teresa, EPL 95, 37002 (2011).
[34] M. Rudolph and J. J. Heremans, Phys. Rev. B **83**, 205410 (2011).
[35] M. Rudolph and J. J. Heremans, Appl. Phys. Lett. **100**, 241601 (2012).
[36] H. Emoto, Y. Ando, E. Shikoh, Y. Fuseya, T. Shinjo and M. Shiraishi, J. Appl. Phys. **115**, 17C507 (2014).
[37] D. Hou, Z. Qiu, K. Harii, Y. Kajiwara, K. Uchida, Y. Fujikawa, H. Nakayama, T. Yoshino, T. An, K. Ando, Xiaofeng Jin and E. Saitoh, Appl. Phys. Lett. **101**, 042403 (2012).
[38] P. Cucka and C.S. Barrett, Acta Crystall. **15**, 865 (1962).
[39] K. Ando, S. Takahashi, J. Ieda, H. Kurebayashi, T. Trypiniotis, C. H. W. Barnes, S. Maekawa and E. Saitoh, Nature Mater. **10**, 655 (2011).
[40] J. C. Rojas-Sánchez, N. Reyren, P. Laczkowski, W. Savero, J. P. Attané, C. Deranlot, M. Jamet, J. M. George, L. Vila and H. Jaffrès, Phys. Rev. Lett. 112, 106602 (2014).
[41] E. E. Krasovskii, F. Starrost, and W. Schattke, Phys. Rev. B **59**, 10504 (1999).

# SUPPLEMENTAL MATERIAL

## I. Resistivity and spin absorption in Bi as a function of temperature.

The resistivity ($\rho$) of Bi as a function of temperature shows the semimetallic behavior expected for Bi, as can be seen in Fig. S1(a). Although the spin absorption ($\eta$) of Bi is reported at 10 and 300 K in the main text, it is also measured at intermediate temperatures [Fig. S1(b)]. The spin diffusion length ($\lambda_{Bi}$) that we would obtain from $\eta$ and Eq. 1 of the main text is plotted as a function of temperature [Fig. S1(c)].

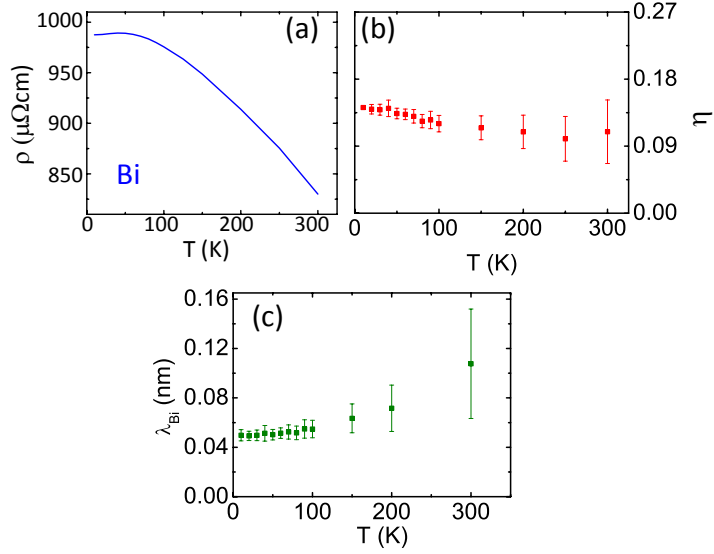

Figure S1. (a) Resistivity of the Bi wire as a function of temperature. (b) Spin absorption parameter, $\eta = \frac{\Delta R_{NL}^{abs}}{\Delta R_{NL}^{ref}}$, as a function of temperature. (c) The spin diffusion length that we would obtain for Bi assuming Eq. 1, as a function of temperature. All measurements correspond to sample 1.

## II. Magnetization rotation of Py

The non-local resistance, $R_{SCC}$, measured in the Bi wire as a result of the IREE should increase with increasing the magnetic field and saturate above the saturation field of the Py injector, following the magnetization rotation of the injector. From the anisotropic magnetoresistance (AMR) measurements of the Py injector [Fig. S2 (a)], we can obtain the angular dependence of the magnetization ($\Psi$) with respect to the injected current using the well known AMR equation: $\rho = \rho_\perp + (\rho_{||} - \rho_\perp)(\cos\psi)^2$ (Fig. S2 (b)). Note that $\Psi$ is the angle between the magnetization ($\vec{M}$) of the Py and the applied charge current ($I_c$), as sketched in the inset of Fig. S2(b).

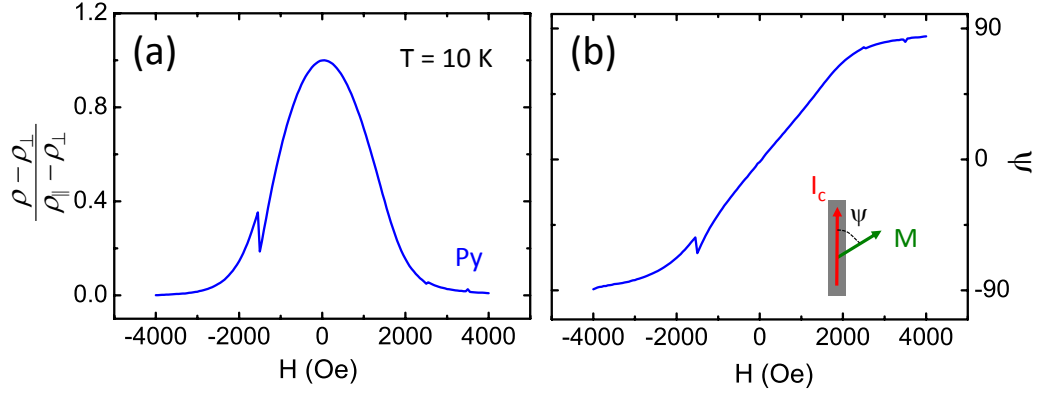

Figure S2. (a) Anisotropic magnetoresistance of Py at 10 K. (b) Angular dependence of the magnetization obtained from : $\rho = \rho_\perp + (\rho_\parallel - \rho_\perp)(\cos\psi)^2$.

### III. Weak antilocalization measurements to extract the spin diffusion length of bulk Bi.

Weak antilocalization (WAL) is a phenomenon by which the resistance of a metal with strong spin-orbit coupling is changed when applying a perpendicular magnetic field at low temperatures [S1]. WAL measurements can be done in 1D, 2D or 3D regimes [S2], and fitting the curve to the corresponding equation allows determining different electronic lengths. The spin-orbit scattering length ($L_{so}$) extracted from WAL measurements is directly related to the spin diffusion length ($\lambda$) in the following way [S3, S4]:

$$\lambda = \frac{\sqrt{3}}{2} L_{so} . \quad (S1)$$

This equation is valid as long as (i) the Elliott-Yafet (EY) mechanism governs the spin scattering and (ii) the Fermi surface is isotropic [S3,S4]. The first statement is fulfilled for the case of bulk Bi as its crystallographic structure shows inversion symmetry and the only mechanism operating in the bulk is therefore EY [S4, S5, S6]. According to the second statement, even if the Bi Fermi surface is not fully isotropic, WAL experiments have also been successfully performed in other non-monovalent metals, such as Pt [S3]. The only implication this may have is that the prefactor that relates $L_{so}$ to $\lambda$ might not be exactly $\sqrt{3}/2$, however it should not be so different from this value. Therefore, Eq. (S1) can be used to estimate the spin diffusion length of bulk Bi from WAL measurements. In this work, we have measured WAL in a Bi thin film, which corresponds to a 2D regime (*i.e.*, the phase coherence length, $L_i$, is larger than the thickness of the films). In this case, a 20-nm-thick film of Bi was e-beam evaporated onto a SiO$_2$ substrate under the same conditions than the Bi used for the lateral spin valves. The measured WAL signal for the 2D regime (Fig. S3) is fitted by the following equation:

$$\Delta G_{sheet}(H) = -\frac{e^2}{2\pi^2\hbar}\left[\Psi\left(\frac{1}{2} + \frac{H_e + H_{so}}{H}\right) - \frac{3}{2}\Psi\left(\frac{1}{2} + \frac{H_e + \frac{4}{3}H_{so}}{H}\right) + \frac{1}{2}\Psi\left(\frac{1}{2} + \frac{H_i}{H}\right)\right], \quad (S2)$$

where $\Delta G_{sheet}$ is the change in conductivity in the Bi film due to WAL, $\Psi$ is the digamma function, and $H_e$, $H_{so}$ and $H_i$ are the elastic, spin-orbit and inelastic scattering fields, respectively. These fields are related to their respective characteristic lengths by:

$$L_{e,so,i} = \sqrt{\frac{\hbar}{4eH_{e,so,i}}}. \quad (S3)$$

By fitting Eq. (S2) to our WAL measurements (see Fig. S3), $L_{so}$=23 nm is obtained and, therefore, $\lambda$=20 nm can be extracted directly. Our previous results indicate that $L_{so}$ is an intrinsic parameter of our Bi films [S7].

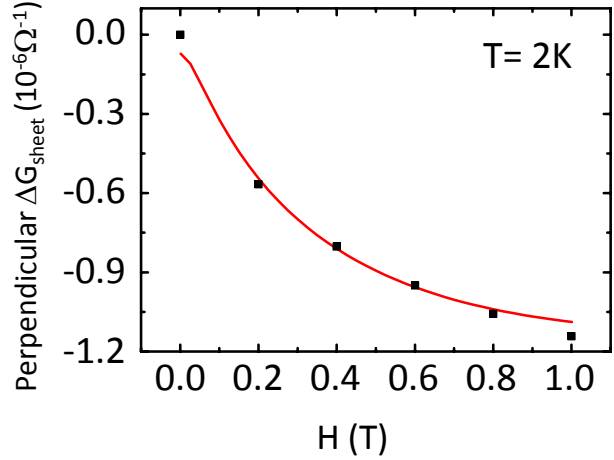

Figure S3. Variation of the sheet conductance for a 20-nm-thick Bi film in a perpendicular magnetic field due to weak antilocalization at 2 K. The red solid line is a fit of the data to Eq. (S2).

## IV. Interface resistance and shunting factor at the Cu/Bi interface as a function of temperature.

The resistance of the Cu/Bi interface, $r_i$, was measured as a function of temperature using a cross configuration (Fig. S4(a) and inset). This resistance is taken into account to obtain the shunting factor, $x$, using a finite elements method (SpinFlow3D software). Note that the shunting factor, plotted in Fig. S4(b) as a function of temperature, is crucial for a proper evaluation of the spin-to-charge conversion.

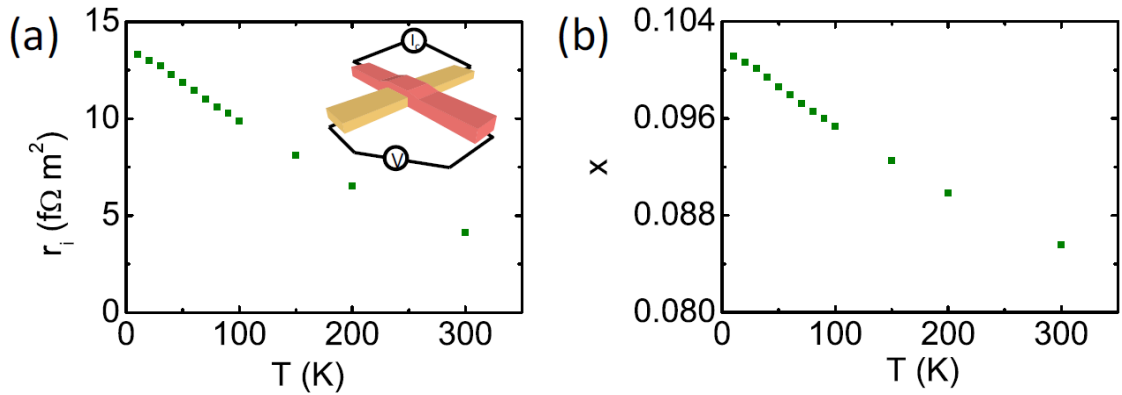

Figure S4. (a) Interface resistance of Cu/Bi as a function of temperature. Inset: Schematic representation of the measurement set up to determine the interface resistance, $r_i$. (b) Shunting factor, $x$, calculated using SpinFlow3D software. All measurements correspond to sample 1.

## V. Calculation of the inverse Spin Hall effect in Bi by considering spin memory loss at the Cu/Bi interface.

In the main text, we consider the possibility that the spin-to-charge conversion arises from the inverse spin Hall effect (ISHE). This would be the case if the spin current diffusing along the Cu channel were absorbed by the bulk Bi, instead of the surface. In such scenario, the spin diffusion length obtained from the SA experiment (which we will call here $\lambda_{Bi}^*$) should be much longer, similar to the lengths obtained from WAL [S7] or spin pumping [S8, S9, S10] measurements (which we assume here to be the bulk value, $\lambda_{Bi}$). Since this is not the case, the observed underestimation could only be compatible with spin absorption in the bulk Bi by assuming a large spin-flip scattering at the interface, leading to spin memory loss (SML) [S11, S12].

Therefore, in order to analyze the effect of SML in our system, we have to study the behavior of the spin currents and the spin accumulation at the interface. A spin accumulation can be quantified by the spin-splitting of the chemical potential $\mu_S = \mu_\uparrow - \mu_\downarrow$, where $\mu_{\uparrow(\downarrow)}$ is the chemical potential for the up (down) spins. The pure spin current density ($j_s$) associated to the diffusion of the spin accumulation, assuming a one-dimensional system, can be expressed as [S13]:

$$j_s = -\frac{1}{e\rho}\nabla\mu_s = -\frac{\lambda}{er_s}\nabla\mu_s, \tag{S4}$$

where $e$ is the electron mass, $\rho$ is the resistivity and $\lambda$ is the spin diffusion length. Note that the spin resistance, defined here as $r_s = \rho\lambda$, quantifies the tendency of the metal to absorb spin currents.

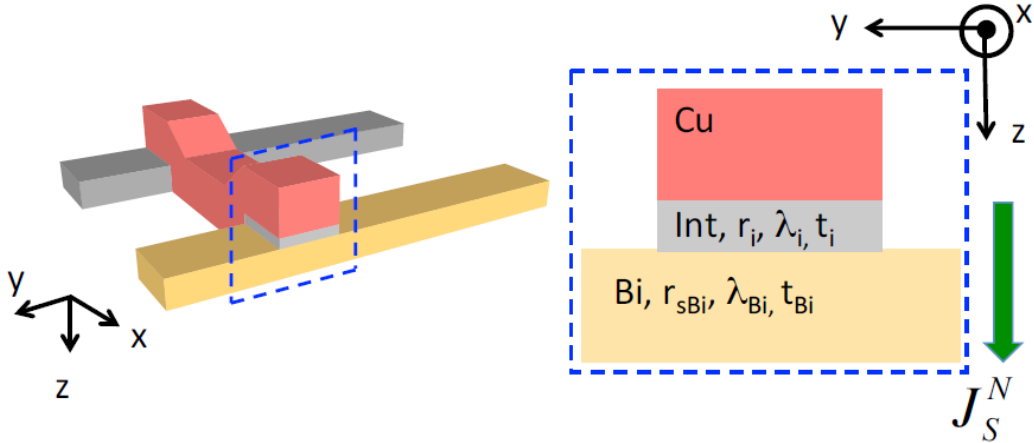

Figure S5. Schematic representation of a device and its transverse cut where the Cu channel, an interfacial layer and the Bi wire are represented. The interfacial layer is inserted to take into account the spin memory loss.

To model our system, let us consider a trilayer formed by Cu|interface|Bi (Fig. S5). When the spin current diffusing along the Cu is absorbed into the Bi wire, the spin current will go through the interface first. Therefore, when defining the spin resistance, we will have to take into account the series resistance ($r_{series}$) of the interface ($r_{si}$) and the Bi layer ($r_{sBi}$). The spin resistance in a series connection is given by [S12, S13]:

$$r_{series} = r_{si} \frac{r_{sBi} \coth\left[\frac{t_{Bi}}{\lambda_{Bi}}\right] \cosh[\delta] + r_{si} \sinh[\delta]}{r_{sBi} \coth\left[\frac{t_{Bi}}{\lambda_{Bi}}\right] \sinh[\delta] + r_{si} \cosh[\delta]} \quad , \tag{S5}$$

where $\delta = t_i/\lambda_i$ is the physical parameter governing the SML, and $t_{Bi,i}$ and $\lambda_{Bi,i}$ are the thicknesses and spin diffusion lengths of Bi and the interface, respectively. The interface spin resistance, $r_{si}$, is related to the interface resistance (Fig. S4(a)), $r_i$, by $r_{si} = r_i/\delta$.

Assuming the SML hypothesis, the spin resistance used in Eq. (1) from the main text that accounts for the spin absorption in the Bi wire should be $r_{series}$. We can thus obtain the SML parameter, $\delta$, by considering $r_{series} = \rho_{Bi}\lambda_{Bi}^*/\tanh[t_{Bi}/\lambda_{Bi}^*]$ and Eq. (S5). The obtained values are $\delta$= 26.93 (4.54) at 10 (300) K.

Once we know $\delta$, we can calculate the spin memory loss parameter, $r_{SML}$, which is the ratio between the spin current absorbed into the Bi wire ($j_s^{Bi}$) and the total spin current coming from the Cu channel ($j_s^{Cu}$):

$$r_{SML} = \frac{j_s^{Bi}}{j_s^{Cu}} = \frac{r_{si}}{r_{sBi} \coth\left[\frac{t_{Bi}}{\lambda_{Bi}}\right] \sinh[\delta] + r_{si} \cosh[\delta]} \quad . \tag{S6}$$

We obtain $r_{SML}$= 7.63×10$^{-15}$ (2.87×10$^{-4}$) at 10 (300) K. This ratio $r_{SML}$ should be taken into account when estimating the ISHE. The spin Hall angle, $\theta_{SH}$, which is the parameter that quantifies the ISHE, can be calculated using the following equation:

$$\theta_{SH} = \frac{1}{r_{SML}} \frac{w_{Bi} t_{Bi}}{x \rho_{Bi} \lambda_{Bi}} \frac{1 - e^{-2t_{Bi}/\lambda_{Bi}}}{\left(1 - e^{-t_{Bi}/\lambda_{Bi}}\right)^2} \left(\frac{I_c}{I_s}\right) \Delta R_{SCC} \quad , \tag{S7}$$

where all the parameters not introduced here are defined in the main text. We obtain $\theta_{SH}$= 3.6×10$^{11}$ (24.7) at 10 (300) K, which are unphysical spin Hall angles (it cannot be larger than 1). Therefore, this rules out the possibility of ISHE in our system.

**VI. Hall measurements for bulk Bi as a function of temperature.**

In order to characterize the electrical properties of Bi, we performed ordinary Hall measurements for a Bi thin film grown in the same conditions as the LSVs and patterned into a Hall bar geometry (see Fig. S6(a)). In these measurements, we obtain the Hall resistivity, $\rho_{Hall}$, by measuring the transverse resistance when applying an out-of-plane magnetic field. The temperature dependence of the $\rho_{Hall}$ exhibits a clear sign change (see Fig. S6(b)), surprisingly close to the experimentally observed change when studying the IREE.

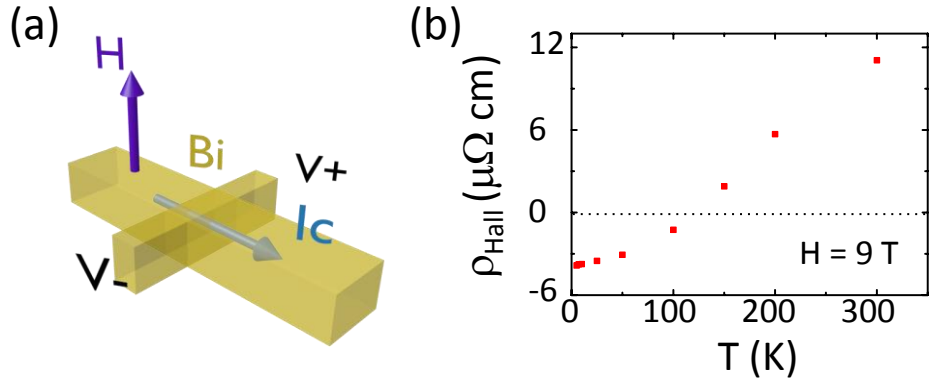

Figure S6. (a) Schematic representation of the measurement set-up to determine the ordinary Hall effect. (b) Temperature evolution of the Hall resistivity, $\rho_{Hall}$, for an external magnetic field of 9T.

However, as the ordinary Hall effect and the IREE are very different phenomena (the former does not depend upon the spin structure, and the latter does) a further theoretical analysis is needed in order to relate both effects, which is beyond the scope of this work.

**References:**

[S1] A. Bergmann, Phys. Rep. **107**, 1 (1984).
[S2] J. Bass and W. P. Pratt, J. Phys.: Cond. Matt. **19**, 183201 (2007)
[S3] Y. Niimi, D. Wei, H. Idzuchi, T. Wakamura, T. Kato and Y. Otani, Phys. Rev. Lett. **110**, 016805 (2013).
[S4] I. Zutic, J. Fabian and S. Das Sarma, Rev. Mod. Phys. **76**, 323 (2004).
[S5] J. Fabian, A. Matos-Abiague, C. Ertler, P. Stano and I. Zutic, Acta Physica Slovaca **57**, 565 (2007).
[S6] M. W. Wu, J. H. Jiang, and M. Q. Weng, Phys. Rep. **493**, 61 (2010).
[S7] S. Sangiao, N. Marcano, J. Fan, L. Morellón, M. R. Ibarra and J. M. De Teresa, EPL **95**, 37002 (2011).
[S8] S. Sangiao, J. M. De Teresa, L. Morellon, I. Lucas, M. C. Martinez-Velarte, and M. Viret, Appl. Phys. Lett. **106**, 172403 (2015).
[S9] H. Emoto, Y. Ando, E. Shikoh, Y. Fuseya, T. Shinjo and M. Shiraishi, J. Appl. Phys. **115**, 17C507 (2014).
[S10] D. Hou, Z. Qiu, K. Harii, Y. Kajiwara, K. Uchida, Y. Fujikawa, H. Nakayama, T. Yoshino, T. An, K. Ando, Xiaofeng Jin and E. Saitoh, Appl. Phys. Lett. **101**, 042403 (2012).
[S11] T. Valet, A. Fert, Phys. Rev. B **48**, 7099 (1993).
[S12] J. C. Rojas-Sánchez, N. Reyren, P. Laczkowski, W. Savero, J. P. Attané, C. Deranlot, M. Jamet, J. M. George, L. Vila and H. Jaffrès, Phys. Rev. Lett. **112**, 106602 (2014).
[S13] T. Kimura, J. Hamrle and Y. Otani, Phys. Rev. B **72**, 014461 (2005).